\documentclass[a4paper,fleqn,usenatbib]{mnras}

\usepackage{newtxtext,newtxmath}

\usepackage[T1]{fontenc}
\usepackage{ae,aecompl}

\usepackage{graphicx}	
\usepackage{amsmath}	
\usepackage{amssymb}	

\usepackage{bm}
\usepackage{tabularx}

\usepackage{soul}
\usepackage[section]{placeins}

\usepackage{color}
\usepackage[dvipsnames]{xcolor}

\usepackage{float}

\title[Turbulent Damping in CR Driven Galactic Winds]{Role of Cosmic Ray Streaming and Turbulent Damping in Driving Galactic Winds}

\author[Holguin et al..]{
F. Holguin,$^{1}$\thanks{E-mail: opaco@umich.edu (FH)}
M. Ruszkowski,$^{1}\thanks{E-mail: mateuszr@umich.edu (MR)}$
A. Lazarian$^{2}\thanks{E-mail: alazarian@facstaff.wisc.edu (AL)}$
R. Farber,$^{1}$\thanks{E-mail: rjfarber@umich.edu (RF)}
and H.-Y. K. Yang$^{3}\thanks{E-mail: hsyang@astro.umd.edu (KY)}$
\\
$^{1}$Department of Astronomy, University of Michigan, Ann Arbor, MI, USA 1\\
$^{2}$Department of Physics and Astronomy, University of Wisconsin-Madison, Madison, WI, USA 2\\
$^{3}$Department of Astronomy and Joint Space Science Institute, University of Maryland-College Park, MD, USA 3}

\date{Accepted XXX. Received YYY; in original form ZZZ}

\pubyear{2018}

\begin{document}
\label{firstpage}
\pagerange{\pageref{firstpage}--\pageref{lastpage}}
\maketitle

\begin{abstract}
Large-scale galactic winds driven by stellar feedback are one phenomenon that influences the dynamical and chemical evolution of a galaxy, redistributing material throughout the circumgalatic medium. Non-thermal feedback from galactic cosmic rays (CRs) --high-energy charged particles accelerated in supernovae and young stars -- can impact the efficiency of wind driving. The streaming instability limits the speed at which they can escape. However, in the presence of turbulence, the streaming instability is subject to suppression that depends on the magnetization of turbulence given by its Alfv\'en Mach number. While previous simulations that relied on a simplified model of CR transport have shown that super-Alfv\'enic streaming of CRs enhances galactic winds, in the present paper we take into account a realistic model of streaming suppression. We perform three-dimensional magnetohydrodynamic simulations of a section of a galactic disk and find that turbulent damping dependent on local magnetization of turbulent interstellar medium (ISM) leads to more spatially extended gas and CR distributions compared to the earlier streaming calculations, and that scale-heights of these distributions increase for stronger turbulence. Our results indicate that the star formation rate increases with the level of turbulence in the ISM. We also find that the instantaneous wind mass loading is sensitive to local streaming physics with the mass loading dropping significantly as the strength of turbulence increases.
\end{abstract}




\section{Introduction}


The baryon-to-halo mass ratio in galaxies is considerably lower than the cosmological average \citep{bell2003}.  At $L_{\ast}$, roughly the Milky Way (MW) luminosity, about 20 percent of baryons are accounted for when matching the observed luminosity to the halo mass function, while at higher or lower luminosities the discrepancy widens \citep{guo2010}. Additionally, absorption lines in background quasars provide evidence for the pollution of the intergalactic medium (IGM) with the products of stellar evolution formed only deep in the galactic potential well, such as dust \citep[e.g.,][]{menardfukugita2012} and metals \citep[e.g.,][]{songaila2001}, up to at least redshift $z = 6$, suggesting that galactic baryons were expelled due to feedback.


At higher luminosities than $L_{\ast}$, feedback from active galactic nuclei (AGN) dominates \citep[e.g.,][]{croton2006}.  For lower luminosities, stellar feedback can drive galactic outflows, pushing and redistributing material, significantly affecting the dynamical and chemical evolution of galaxies \citep{larson1974, whiterees1978, duboisteyssier2010}. Indeed, galactic winds have been observed in galaxies that have had recent and significant star formation \citep{veilleux2005}, driving gas out at a rate of 0.01 to 10 times the star formation rate (SFR) \citep{blandhawthorn2007}.

The stellar feedback which drives winds is likely the result of several mechanisms combining in a non-linear manner \citep[e.g.,][]{agertz2013}. A detailed understanding of the exact mechanisms and their complex interactions remains uncertain, as many processes operate below the grid scale of simulations in galactic and cosmological simulations \cite[][]{somervilledave2015}. 

%
Mechanisms used to explain winds are thermal \citep{chevalierclegg1985, joung2009} and momentum \citep{kim2016} feedback from supernovae (SN), as well as radiation pressure from massive stars \citep{murray2005, murray2011, hopkins2012}. Galactic cosmic rays (CRs), originating from shock acceleration in SN remnants \citep[see][]{bykov2018} and winds from massive stars \citep[see][]{bykov2014} can also play a significant role in launching galactic winds. In the MW, the CR energy density is in rough equipartition with the turbulent and magnetic field energy densities \citep[e.g.,][]{boularescox1990}. Additionally, Fermi $\gamma$-ray observations of starburst galaxies M82 and NGC 253 suggest CR energy densities two orders of magnitude above the MW values \citep{paglione2012, yoasthull2013}. These two findings hint at the importance of CRs in the evolution of galaxies.
Theoretical considerations suggest that CRs can play important role in driving gas in galactic winds \citep{everett2008, breitschwerdt1991}. Three-dimensional hydrodynamical \citep{uhlig2012, booth2013, salembryan2014}
 and magneto-hydrodynamical (MHD) \citep{hanasz2013, girichidis2016, pakmor2016} simulations have demonstrated that CRs indeed influence the generation of global outflows and the local structure of the interstellar medium (ISM).  The exact properties of the simulated outflows depend sensitively on how CR transport is modeled \citep{simpson2016, ruszkowski2017,farber2018}.  \\
\indent 
In the self-confinement model of CR transport, CRs propagating in one direction along the magnetic field in the galaxy generate Alfv\'en waves that scatter CRs back, thus amplifying the waves. This process is called the streaming instability and in the absence of wave dissipation it was shown to reduce the CR bulk streaming speed $u_s$ (relative to the gas) to the Alfv\'en speed $u_A$ \citep[see][]{kulsrudpearce1969}. This effectively couples the plasmas with the CRs. In terms of galactic winds this means that the flux of CRs can transfer its momentum to the wind material. 

If the dissipation of Alfv\'en waves is present, the streaming instability can still be present and the coupling of the waves and the wind is decreased. Historically, the damping of the Alfv\'en waves in the context of streaming instability suppression is associated with the ion-neutral linear damping process \citep{kulsrudpearce1969}. This process is not efficient for the highly ionized matter expected to form galactic winds. However, it was noted in \citet{yanlazarian2002} that the streaming instability can be suppressed by turbulence. \citet{farmergoldreich2004} proposed a model for trans-Aflv\'enic strong MHD turbulence, corresponding to the Alfv\'en Mach number $M_A=u_L/u_A = 1$, where $u_L$ is the injection velocity at the turbulence injection scale $L$. This study was generalized for the arbitrary $M_A$ in \citet{lazarian2016}, where it was shown that the damping significantly changes with $M_A$. Moreover, the latter study showed that the scaling for the dependences of the damping of the streaming instability for $M_A<1$ is different for high energy CRs that induce waves that are non-linear damped by the {\it weak} Alfvenic turbulence \citep[see][]{lazarianvishniac1999, galtier2000} that spans the range from $LM_A^2$ to $L$ and the lower energy cosmic rays that induce waves that are non-linearly damped by the {\it strong} MHD turbulence existing at the scales less than $LM_A^2$. Note, that the terms {\it weak} and {\it strong} turbulence do not reflect the amplitude of Alfv\'enic perturbations, but the strength of non-linear interactions \citep[see][]{brandenburglazarian2013}. 

Observations of the MW suggest that the case of $M_A<1$ is the most appropriate for the turbulence at high galactic latitudes corresponding to the action of the galactic wind \cite[see][]{kandel2018}. 
\noindent The magnetically dominated, i.e. low $\beta$ \footnote{$\beta$ is the ratio of the gaseous pressure to the magnetic pressure.}  media is also expected for the galactic wind environment. Therefore we do not consider the non-linear Landau damping \citep{zweibel2013} that may be important for the damping of Alfven waves in high $\beta$ media. The turbulent damping of the streaming instability is a robust process that depends only on the turbulence properties and does not depend on plasma $\beta$. Our present study is focused on studying the consequences of this process for the generation and the evolution of galactic winds and resulting galactic properties. 

The important earlier work that considered the effects of the launching of the galactic winds with CRs and taking into account the effects of the streaming instability is \citet{ruszkowski2017}. There, a simple parameterization $u_s = f u_A$ was considered for the CR streaming speed in three-dimensional MHD simulations of an isolated galaxy ($f$ was assumed to be constant). In these simulations, the CR streaming generally enhances the efficiency of galactic wind driving, as the CRs can escape from dense regions and interact with more tenuous gas that is easier to accelerate.
As the efficiency of wind coupling with CRs is determined by the efficiency of the turbulent damping of the streaming instability, it is essential to properly model this process. In implementing a more physically motivated model compared to that adopted by \citet{ruszkowski2017}, we use the model of turbulent damping in \cite{lazarian2016} and provide the more accurate description of the turbulent damping of Alfv\'en waves. We calculate the resulting streaming speed that depends on the local properties of the ISM and halo.
We perform three-dimensional MHD simulations of a section of a MW-like galactic disk in order to investigate the effects of locally-determined CR streaming controlled by the turbulent structure of the ISM.  We include magnetic fields, radiative cooling, self-gravity, and stellar feedback (star formation and SN, with thermal and CR injection). In Section 2, we describe the treatment of numerical methods and physical models, while in Section 3 we discuss results, with conclusions in Section 4.


\section{Methods}
We run simulations with the adaptive mesh refinement MHD code FLASH 4.2 \citep{fryxell2000, dubey2008} using a directionally unsplit staggered mesh (USM) solver \citep{leedeane2009, lee2013}, including CR physics \citep{yang2012, ruszkowski2017, farber2018}, in an elongated box of dimensions $2 \times 2 \times 40$ $\textrm{kpc}^3$.

We solve the MHD equations with a two-fluid model \citep{salembryan2014, ruszkowski2017}, including both thermal gas and ultra-relativistic CR fluid (composed of protons) characterized by adiabatic indices $\gamma = 5/3$ and $\gamma_{\rm cr} = 4/3 $, respectively. We use a mean CR Lorentz factor $\gamma_{\textrm{rel}} = 3$, and  a slope $n = 4.5$ for the CR distribution function in momentum, which are typical values for galactic CRs.

We include CR advection, dynamical coupling between CRs and thermal gas, CR streaming along the magnetic field lines and the associated heating of gas by CRs, gas self-gravity, radiative cooling, star formation, and evolve the following equations:

\begin{align}
\frac{\partial \rho}{\partial t} + \bm{\nabla} \cdot ( \rho \bm{u}_g )   =  -\dot{m}_{\textrm{form}} + f_{\ast} \dot{m}_{\textrm{feed}}
\label{eom1}
\end{align}

\begin{equation}
\frac{\partial \rho \bm u_g}{\partial t} + \bm{\nabla} \cdot \left(  \rho \bm{u}_g \bm{u}_g - \frac{\bm{B}\bm{B}}{4\pi} \right) + \bm{\nabla} p_{\textrm{tot}} = \rho \bm{g} + \dot{p}_{\rm SN}
\label{eom2}
\end{equation}

\begin{equation}
\frac{\partial \bm{B}}{\partial t} - \bm{\nabla} \times (\bm{u}_g \times \bm{B} ) = 0
\label{eom3}
\end{equation}

\begin{equation}
\begin{split}
\frac{\partial e}{\partial t} + & \bm{\nabla} \cdot \left[ (e + p_{\textrm{tot}}) \bm{u}_g - \frac{\bm{B}(\bm{B} \cdot \bm{u}_g  )}{4\pi}    \right]  = \rho \bm{u}_g \cdot \bm{g}  \\
&- \nabla \cdot \bm{F}_{\rm cr}  - C + H_{\rm SN}
\end{split}
\label{eom4}
\end{equation}

\begin{equation}
\begin{split}
\frac{\partial e_{\rm cr} }{\partial t} + \bm{\nabla} \cdot(e_{\rm cr} \bm{u}_g ) = & -p_{\rm cr}  \nabla \cdot \bm{u}_g - H_{\rm cr} + H_{\rm SN} \\
& - \nabla \cdot \bm{F}_{\rm cr} 
\end{split}
\label{eom5}
\end{equation}

\begin{equation}
\Delta \phi = 4\pi G \rho_b
\label{eom6}
\end{equation}

\noindent where $\rho$ is the gas density, $\rho_b$ is the total baryon density including both the gas and stars, $\dot{m}_{\textrm{form}}$ is the density sink from stellar population particle formation, $f_{\ast} \dot{m}_{\textrm{feed}}$ represents the gas density source from stellar feedback (see Section \ref{sfeed}), $\bm{u}_g$ is the gas velocity, $\bm{B}$ is the magnetic field, $G$ is the gravitational constant, $\phi$ is the gas gravitational potential, $\bm{g} = - \bm{\nabla} \phi + \bm{g}_{\textrm{NFW}}$ is the gravitational acceleration (the sum of gas self-gravity, stellar particle, and halo dark matter contributions to the gravitational acceleration, described in Section \ref{grav}) where $\bm{g}_{\textrm{NFW}}$ is the gravity from the Navarro-Frenk-White (NFW) potential, $p_{\textrm{tot}}$ is the sum of gas ($p_{\rm th}$), magnetic, and CR ($p_{\rm cr}$) pressures, $\dot{p}_{\rm SN}$ is the momentum injection due to stellar winds and SN.  Furthermore, $e = \rho \bm{u}^2_g + e_g + e_{\rm cr} + B^2 / 8\pi$ is the total energy density per volume (the sum of gas, CR, and magnetic components, respectively), $C$ is the radiative cooling rate per unit volume, and $H_{\rm SN}$ is the supernova heating rate per volume.
CR advection and coupling to the gas are included using the same methods as in \citet{ruszkowski2017} \citep[see][]{yang2012, yang2013, sharma2009} with the CR streaming flux denoted by $\bm{F}_{\rm cr}$ and an associated CR heating of the gas denoted by $H_{\rm cr}$. The streaming flux is $\bm{F}_{\rm cr} = (e_{\rm cr} + p_{\rm cr}) \bm{u}_s$, with the streaming speed along the magnetic field down the CR gradient $\bm{u}_s \propto -\textrm{sgn}(\hat{\bm{b}} \cdot \nabla e_{\rm cr} ) \simeq \textrm{tanh}(h_c \hat{\bm{b}} \cdot \nabla e_{\rm cr} / e_{\rm cr}) $, where $\hat{\bm{b}}$ is the magnetic direction vector. When damping processes are included, we write $\bm{u}_s = f \bm{u}_A$, where f is a function of local gas properties (see Section 2.4). The regularization parameter $h_c = 10$ kpc helps avoid prohibitively small time-steps due to discontinuities in $\bm{F}_{\rm cr}$ near extrema of the CR energy density distribution \citep{sharma2009}.  The streaming speed is limited to 200 km/s for computational efficiency. We have tested higher ceilings finding no significant change in the results. Additionally, we sub-cycle four times over the CR streaming term to further accelerate computations.

\subsection{Gravity} \label{grav}

The gravitational acceleration has contributions from the gravity of gas, stellar particles, and dark matter halo vertical component. We do not include a pre-existing stellar potential, instead we allow the gravitational contribution from stars to be set up by the stellar particles. For the self-gravity of baryons, we solve the Poisson equation with the Barnes-Hut tree solver \citep{barneshut1986} implemented by \citet{wunsch2018}. We also include the gravitational contribution from the overall dark matter halo \citep{navarro1997}.  Since the domain is a thin slice of a galaxy, we only use the vertical component of gravity

\begin{equation}
g_{\textrm{NFW}} (z) = - \frac{G M_{200}}{|z|^3} \frac{\textrm{ln}(1+ x)    - x/(1+x)  }{\textrm{ln}(1+c) - c/(1+c)}
\end{equation}

\noindent where $G$ is the gravitational constant, $M_{200}$ is the halo virial mass, $z$ is the height above the mid-plane, $x = |z| c / r_{200}$, $c$ is the halo concentration parameter, and $r_{200}$ is the virial radius. Table \ref{table1} summarizes the parameters we use.

\subsection{Radiative cooling}

We use the Townsend cooling method \citep{townsend2009,zhu2017} implemented as in \citet{farber2018}.  The cooling function $\Lambda(T)$ is a piecewise power law with a floor temperature of 300 K \citep{rosenbregman1995} given in units of erg $\textrm{cm}^3 \textrm{s}^{-1}$ by

\begin{equation}
\Lambda(T) = \left\{\begin{aligned}  
    &0                                 &&\mbox{if                            $T$ $<$             300} \\
    &2.2380\times 10^{-32}T^{2.0  } &&\mbox{if               300 $\leq$ $T$ $<$            2000} \\
    &1.0012 \times 10^{-30}T^{1.5} &&\mbox{if              2000 $\leq$ $T$ $<$            8000} \\
    &4.6240 \times 10^{-36}T^{2.867} &&\mbox{if              8000 $\leq$ $T$ $<          10^{5}$} \\
    &1.7800 \times 10^{-18}T^{-0.65} &&\mbox{if          $10^{5}$ $\leq$ $T$ $< 4 \times 10^{7}$} \\
    &3.2217 \times 10^{-27}T^{0.5} &&\mbox{if $4 \times 10^{7}$ $\leq$ $T$, }\end{aligned}\right. 
\label{cooling}
\end{equation}

\noindent where $T$ is the gas temperature in K. This cooling function is an approximation to the radiative cooling functions in \citet{dalgarnomccray1972} and \citet{raymond1976}, accurate for a gas of solar abundance that is completely ionized gas at $T = 8000$ K.  The Townsend scheme does not impose restrictions on the cooling time step.

\subsection{Star formation and feedback} \label{sfeed}

Star formation follows the approach of \citet{cenostriker1992} \cite[see also][]{taskerbryan2006, bryan2014, salembryan2014, li2015}, where star formation occurs when all of the following conditions are met: (i) gas density exceeds $1.67 \times 10^{-23} \textrm{g cm}^{-3}$ \citep{gnedinkravtsov2011, agertz2013}; (ii) the cell mass exceeds the local Jeans mass; (iii) $\nabla \cdot \bm{u}_g < 0$; (iv) gas temperature reaches the floor of the cooling function or the cooling time becomes shorter than the dynamical time $t_{\textrm{dyn}} = \sqrt{3\pi / (32G\rho_b)}$.  When these conditions are satisfied, a stellar population particle is formed instantaneously at a random position in the cell, with the same velocity as the gas, and mass $m_{\ast} = \epsilon_{\textrm{SF}} (dt/t_{\textrm{dyn}}) \rho dx^3 $, where $\epsilon_{\textrm{SF}} = 0.05$ is the star formation efficiency, $dt$ is the timestep, and  $dx$ is the local cell size. There is a corresponding removal of gas mass from the surrounding cell.

In order to keep the number of particles manageably small we set a minimum particle mass $m_{*,\textrm{min}} = 10^5 M_{\odot}$. We still permit particles with $m_{\ast} < m_{*,\textrm{min}}$  to form; their masses are given by $m_{\ast} = 0.8 \rho dx^3$ forming with a probability $m_{\ast}/m_{*,\textrm{min}}$.

We include stellar feedback from winds and SN by adding gas mass, thermal energy, and CR energy into the cell surrounding a particle. For this feedback, the assumed stellar mass within a particle is not modeled as instantaneous, as stars form and evolve over time. The stellar mass increases at a rate of $\dot{m} = m_{\ast}(\Delta t /\tau^2) \ \textrm{exp}(-\Delta t/ \tau)$, where $\Delta t$ is the time since formation of the particle, and $\tau$ =  max($t_{\textrm{dyn}}$, 10 Myr). Gas mass is added at a rate $ f_{*} \dot{m}_{\textrm{feed}} = \ f_{*} \dot{m} \ $ into the cell surrounding a stellar population particle.  The injected gas has a velocity equal to that of the source particle, thermal energy equal to $ (1-f_{\textrm{cr}} ) \epsilon_{\textrm{SN}} \dot{m} c^2$, and CR energy equal to $f_{\textrm{cr}} \epsilon_{\textrm{SN}} \dot{m} c^2$, where $f_{\textrm{cr}}$ is the fraction of total SN energy given to CRs.  We assume $f_{*} = 0.25$ for the fraction of returned mass from the star to the ISM and $\epsilon_{\textrm{SN}} = 10^{51}$ erg/($\textrm{M}_{sf} c^2)$ for the energy injected by one supernova per $\textrm{M}_{\textrm{sf}} = 100 M_{\odot}$ \citep{guedes2011, hanasz2013} of mass of a newly formed stellar population particle, corresponding to a \citet{kroupa2001} initial mass function. The mass of the stellar population particle is reduced appropriately after the gas injection into the ISM.  The parameter choices are summarized in Table \ref{table1}.

\begin{table}
  \caption{Model parameters}
  \begin{center}  
  	\leavevmode
    \begin{tabularx}{0.45\textwidth}{Xl}
    \hline\hline
  Halo                  \\ \hline 
   $M_{200}^{(1)}$ &   $10^{12}$M$_{\odot}$      \\
   $c^{(2)}$ &  12                   \\ \hline
   Disk                 \\ \hline
    $\rho_{o}^{(3)}$ & 5.24 $\times 10^{-24}$ g\ cm$^{-3}$ \\
    $z_{o}^{(4)}$ & 0.325 kpc \\
   $\Sigma_{o}^{(5)}$ & 100 M$_{\odot}$\ pc$^{-2}$ \\
	$T_{o}^{(6)}$    & 10$^{4}$ K \\
	$B_{o}^{(7)} = B_{o,x}$  & 3 $\mu$G \\   \hline   
Star Formation  \\ \hline
$n_{\rm thresh}^{(8)}$ & 10 cm$^{-3}$ \\
$T_{\rm floor}^{(9)}$ & 300 K \\
$m_{*,{\rm min}}^{(10)}$ & 10$^{5}$ M$_{\odot}$ \\
$\epsilon_{\rm SF}^{(11)}$ & 0.05 \\ \hline
Stellar Feedback  \\ \hline
$f_{*}^{(12)}$ & 0.25 \\
$f_{cr}^{(13)}$ & 0.1 \\
$\epsilon_{\rm SN}^{(14)}$ & 10$^{51}$erg/(M$_{\rm sf}$c$^{2}$) \\
$M_{\rm sf}^{(15)}$ & 100 M$_{\odot}$ \\ \hline

\end{tabularx}
\label{table1}
\end{center}

\textbf{Notes.} From top to bottom the rows contain: (1) halo mass; (2) concentration parameter; (3) initial midplane density; (4) initial scale height of the gas disk; (5) initial gas surface density; (6) initial temperature; (7) initial magnetic field strength; (8) gas density threshold for star formation; (9) floor temperature; (10) minimum stellar population particle mass; (11) star formation efficiency; (12) fraction of stellar mass returned to the ISM; (13) fraction of supernova energy bestowed unto CRs; (14) SN energy per rest mass energy of newly formed stars; (15) rest mass energy of newly formed stars per SN.

\end{table}
\subsection{Cosmic ray streaming}

In the self-confinement model of CR transport through the ISM, CRs stream along magnetic field lines, exciting Alfv\'en waves due to the streaming instability, subsequently limiting the CR streaming speed to the Alfv\'en speed \citep{zweibel2013}.

As relativistic, charged particles (usually protons), CRs gyrate around a local magnetic field line at the frequency $\Omega_{0} / \gamma_{\textrm{rel}}$ and gyro-radius $r_{L} = \gamma_{\textrm{rel}} c / \Omega_0 $, where $c$ is the speed of light, $\gamma_{\textrm{rel}}$ is the Lorentz factor, $\Omega_0 = eB/m_p c$ is the non-relativistic cyclotron frequency, and $m_{p}$ and $e$ are the proton mass and charge, respectively. CRs strongly interact with Alfv\'en waves when the resonance condition  $k_{||} = 1/(\mu r_{L})$ is met; the Alfv\'en parallel wavevector $k_{||}$ (with respect to the local magnetic field) is of order the inverse of the CR gyro-radius projected onto the plane of the wave by the cosine of the pitch angle $\mu$. At smaller gyro-radii, the local magnetic field does not change much over a CR orbit reducing the interaction, and at larger gyro-radii, the CR samples a large enough spatial region that the effects of the fluctuating field cancel out over the orbit.

Alfv\'en waves are amplified by resonant scattering of CRs at a rate shown by \citet{kulsrudpearce1969} and  \citet{wentzel1974}

\begin{equation}
\Gamma_g \approx \frac{\pi}{6} \Omega_0 \frac{n_{\rm cr} (> \gamma_{\textrm{rel}} )}{n_i}  \left(\frac{u_s}{u_A} - 1 \right),
\end{equation}

\noindent where $n_{\rm cr}({>}\gamma_{\textrm{rel}})$ is the number density of CRs with sufficiently large gyro-radii (directly dependent on $\gamma_{\textrm{rel}}$ and also the CR energy) to be resonant with the Alfv\'en wave, $n_i$ is the ion number density, and $u_{A}$ is the Alfv\'en speed.  CRs with larger gyro-radii can still be resonant as the projection of the orbit to the plane of the wave can meet the resonance condition. Amplification of Alfv\'en waves occurs until the CRs become isotropic in the wave frame and stream at $u_A$. In addition to growth, Alfv\'en waves experience damping by various mechanisms, in particular by ion-neutral friction, non-linear Landau damping, or turbulent damping.  As a result of this damping, CR motion will be super-Alfv\'enic to a degree depending on the damping rate.

Given that the galactic astrophysical environment is magnetized, turbulent, and significantly ionized \citep{mckeeostriker2007, sharma2009, brandenburglazarian2013}, we consider the effects of MHD turbulence stirred up by SN feedback. This was originally suggested by \citet{yanlazarian2002} and quantified in \citet{farmergoldreich2004} using the strong, incompressible MHD cascade \citep{goldreichsridhar1995}.

As Alfv\'en waves pass through turbulent eddies, they are irreversibly distorted. The turbulent eddies are anisotropic and aligned with the magnetic field \citep[e.g.,][]{higdon1984}. As a result of this anisotropy, the Alfv\'en waves that experience the least amount of distortion are those with wave vector parallel to the local magnetic field, which is exactly the case for the waves generated by the streaming instability.  


The damping rates depend on the properties of turbulence that exists in the environment, which is characterized by the turbulent Alfv\'en Mach number $M_A = \frac{\sigma}{u_A} $, where $\sigma$ is the gas velocity dispersion, and the inertial range of the turbulence, depending on the ratio of the CR gyro-orbit to the injection length scale. Turbulence can be sub-Alfv\'enic, $M_A <1$, and either strong or weak, depending on whether $r_L/L$ is greater or less than $M_A^4$ respectively. $L$ is the length scale at which turbulence is driven. Turbulence can also be super-Alfv\'enic, $M_A > 1$, and either strong or hydro-like, depending on whether $r_L/L$ is greater or less than $M_A^{-3}$ respectively. \citet{lazarian2016} provides a general study of the Alfv\'en wave damping rates $\Gamma_d$ for each regime of turbulence. The results are summarized in Table \ref{regimetable}.

\begingroup
\setlength{\tabcolsep}{10pt} 
\renewcommand{\arraystretch}{2} 

\begin{table*}
\centering
\caption{Summary of CR streaming speed boost $f-1$ above Alfv\'enic streaming for four different regimes of MHD turbulence \citep{lazarian2016} with given inertial range, where $M_A$ is the turbulent Mach number, $r_L$ is the CR gyro-radius, and $L$ is the turbulence injection scale. The ratio $l_{\textrm{min}}/L << 1$, where the $l_{\textrm{min}}$ is the thermal ion gyro-radius.}
\begin{tabular}{l l l l l }
\hline\hline
Turbulence & $M_{A}$       & Inertial Range                                                      & $\Gamma_d$                                            & $f-1$                                                                                                \\ \hline
Weak       & \textless1    & $M_A^4 < \frac{r_L}{L} < M_A$                                       &  $\frac{u_A M_{A}^{8/3}}{r_L^{2/3} L^{1/3}}$ & $ \frac{u_A}{\Omega_0} \frac{n_{i}}{n_{\rm cr}}\frac{M_{A}^{8/3}}{r_L^{2/3} L^{1/3} }$ \\ \hline
Strong     & \textless1    & $\left(\frac{l_{\textrm{min}}}{L}\right)^{4/3} < \frac{r_L}{L} < M_A^4$        & $\frac{u_A M_{A}^{2}}{r_L^{1/2} L^{1/2}}$   & $\frac{u_A}{\Omega_0} \frac{n_{i}}{n_{\textrm{cr}}}\frac{M_{A}^{2}}{r_L^{1/2} L^{1/2} }$    \\ \hline
Strong     & \textgreater1 & $\left(\frac{l_{\textrm{min}}}{L}\right)^{4/3} M_A < \frac{r_L}{L} < M_A^{-3}$ & $\frac{u_{A} M_{A}^{3/2}}{r_L^{1/2} L^{1/2}}$ & $ \frac{u_A}{\Omega_0} \frac{n_{i}}{n_{\rm cr}}\frac{M_{A}^{3/2}}{r_L^{1/2} L^{1/2} }$ \\ \hline
Hydro      & \textgreater1 & $M_A^{-3} < \frac{r_L}{L} < 1$                                      & $\frac{u_A M_{A}^{3/2}}{r_L^{2/3} L^{1/3}}$ & $ \frac{u_A}{\Omega_0} \frac{n_{i}}{n_{\rm cr}}\frac{M_{A}^{3/2}}{r_L^{2/3} L^{1/3} }$ \\ \hline
\end{tabular}

\label{regimetable}
\end{table*}
\endgroup

Following \citet{wiener2013} and \citet{ruszkowski2017}, we can parameterize the CR transport speed by balancing the Alfv\'en wave growth and turbulent damping rates.  As an example, we derive the CR streaming speed assuming the weak, sub-Alfv\'enic turbulent damping rate $\Gamma_{d,\textrm{weak}}$ and compare it to the streaming instability growth rate



\begin{equation}
\begin{split}
\Gamma_g \ = \ & \Gamma_{d,\textrm{weak}} \\
\frac{\pi}{6} \Omega_0 \frac{n_{\rm cr} (> \gamma_{\textrm{rel}} )}{n_i}  \left(\frac{u_{s}}{u_{A}} - 1 \right) \ = \ & \frac{M_{A}^{8/3}}{r_{L}^{2/3} L^{1/3}}  u_{A}
\end{split}
\end{equation}

\noindent we then solve for the streaming speed $u_{s}$

\begin{equation}
u_{s} = u_{A}  \left(1 + \frac{u_A}{\Omega_0} \frac{n_{i}}{n_{\textrm{cr}} \left(> \gamma_{\textrm{rel}} \right)}\frac{M_{A}^{8/3}}{r_L^{2/3} L^{1/3}}\right).
\end{equation}

\noindent We set the length scale $L = 10$ pc \citep{iacobelli2013}.  A similar analysis using the damping rates appropriate for the other turbulence regimes yields the remaining three expressions for the streaming speed. In general, we write

\begin{equation}
u_{\textrm{s}} = u_{A} f(n_i , n_{\rm cr}, B, \sigma), 
\label{crss}
\end{equation}

\noindent where the proportionality $f$ is a function of ion and CR number densities, $n_i$ and $n_{\rm cr}$, magnetic field strength $B$, and velocity dispersion $\sigma$. 


The streaming speed boost above the Alfv\'en speed ($f - 1$) is listed in the last column in Table \ref{regimetable}.  The expressions for the boost show expected dependencies on the environment. The streaming boost is proportional to $n_i$ and $\sigma$ and inversely proportional to $B$ and $n_{\rm cr}$. For higher ion density or velocity dispersion, all other parameters fixed, stronger turbulence can more efficiently damp Alfv\'en waves, leading to faster CR propagation. On the other hand, a higher magnetic field strength results in higher growth rate of Alfv\'en waves due to the streaming instability, trapping CRs more effectively and thus reducing their effective speed. Similarly, a greater density of CRs generates more Alfv\'en waves, also trapping CRs more effectively and slowing down CR propagation. In our simulations we use a constant velocity dispersion. The assumption of constant velocity dispersion is consistent with the decay of the turbulence strength from central star formation regions to the halo \citep{stone1998} as the turbulent Alfv\'en Mach number, $M_A$, still decreases with height above the midplane because the Alfv\'en speed increases. We compare results of simulations with two different gas velocity dispersion values $\sigma = 5$ km/s and 10 km/s, which are representative of turbulence in the disk, as the turbulent structure near the disk (where the wind is launched) will most likely dictate the resulting galactic wind structure and evolution \cite[][]{kimostriker2018}. Therefore, we do not expect the results to change significantly if instead of considering constant $\sigma$, we use a $\sigma$ profile declining with distance from the midplane. We choose constant $\sigma$ also because quantifying $\sigma$ on-the-fly in the simulation would pose additional challenges associated with averaging velocities over finite volumes.

CRs experience an effective drag force as they propagate down the pressure gradient $\nabla p_{\rm cr}$ and scatter off MHD waves, leading to heating of gas. The collisionless heating term $H_{\rm cr} = |\bm{u}_A \cdot  \nabla p_{\rm cr}|$ depends on the Alfv\'en speed $u_A$ and not the potentially super-Alfv\'enic streaming speed $u_s$ because the transfer of energy from CRs to the gas is only due to the portion of streaming caused by MHD waves \citep[see the Appendix in][]{ruszkowski2017}. We do not include loss of CR energy due to hadronic or Coulomb losses (collisional heating of gas) in the $H_{\rm cr}$ term.

\subsection{Simulation setup}

We simulate a section of a galactic disk in an box of dimensions $2 \times 2 \times 40$ $\textrm{kpc}^3$, elongated in the direction $z$ above the mid-plane. Previous work has shown the importance of including sufficient height in these types of slab simulations in establishing a realistic temperature distribution in the halo \citep{hill2012}.  We choose the vertical extent of the box to be equal to 40 kpc.  A box of such height is sufficiently extended to limit the above-mentioned problem, while not compromising feasibility of our computations.

We use periodic boundary conditions for the box sides perpendicular to the disk plane and diode boundary conditions for those parallel to the disk plane.  The diode boundary conditions do not allow in-fall back into the box \citep[e.g.,][]{sur2016}. We do not include the effects of differential rotation due to large scale galactic motion in order to simplify the simulation and focus on feedback processes. We use static mesh refinement that varies according to height $|z|$ above the mid-plane, achieving a maximum resolution of 31.25 pc in the disk for $|z| < 2 \ \textrm{kpc}$, and progressively coarser resolution of 62.5 pc for $2 \ \textrm{kpc} < |z| < 5 \ \textrm{kpc}$, and a minimum resolution of 125 pc elsewhere in the halo.
 
We initialize the simulation with the vertical equilibrium density solution for a stratified, isothermal self-gravitating system \citep{spitzer1942, salembryan2014} within a stratified box model \citep[e.g.,][]{deavillez2007,walch2015,farber2018} as follows

\begin{equation}
\rho(z) = \begin{cases} \rho_{0} \hspace{0.1cm}\mathrm{sech}^{2}\left( \frac{z}{2 z_{0}} \right) \ \ \mbox{ $\rho(z) > \rho_{\textrm{halo}}$} \\
                        \rho_{\textrm{halo}}                                       \mbox{\hspace{1.75 cm}otherwise,} \end{cases}
\label{spitzer}
\end{equation}

\noindent where $\rho_0$ is the initial mid-plane density and $z_0$ is the vertical scale height. We can define the initial disk surface gas density as $\Sigma_0 = \int_{-20 \textrm{kpc}}^{20 \textrm{kpc}} \ \rho (z) \textrm{d}z$.  $\rho_{\textrm{halo}} = 1.0 \times 10^{-28} \textrm{g/cm}^3$ is the initial density of the halo. 

The parameters that we chose for the spatial distribution of the gas correspond to total gas surface density of around 100 $100 \ M_{\odot}/\rm{pc}^2$. This value of the surface density is computed by averaging within a radius of 10 kpc the isothermal self-gravitating solution described by Eq. \ref{spitzer}, assuming MW-type baryon fraction. The initial gas surface density is on the order of the average gas surface density at a radius of around $r=8$ kpc. 

The overall gravitational potential is comparable to that of other works. When most of the stellar particles have formed, the stellar particle surface density in the simulations approaches about 20 M$_{\odot}$\ pc$^{-2}$, which is comparable to the pre-existing stellar surface density of 30 M$_{\odot}$\ pc$^{-2}$ assumed by \cite{walch2015}. However, \cite{walch2015} assumes an initial gas surface density of 10 M$_{\odot}$\ pc$^{-2}$, so the gravitational potential due to gas is lower in our simulations.

The magnetic field is initially oriented along a horizontal direction and its magnitude follows the density distribution such that $B(z) \propto \rho(z)^{2/3}$ with the mid-plane value $B_0 \approx 3 \mu G$. We initialize the simulation with a constant temperature $T= 10^4$ K. The setup is initially out of thermal equilibrium. At the beginning of the simulation, the gas distribution will collapse due to radiative cooling, and begin the formation of stellar population particles. Subsequent stellar feedback will suppress star formation at about 50 Myr in the simulation. Some previous works such as \citet{kimostriker2017} initially drive turbulence in the disk in order to achieve convergence in the simulation more quickly. We choose not to include initial artificial pressure support. The resulting steady-state properties are insensitive to the choices for initial artificial pressure support \citep{kimostriker2017}.

In the Alfv\'en wave damping formualae, we assume the gas is completely ionized so $n_i = n_{\rm{gas}}$, where $n_{\rm{gas}} = \rho / m_p$ is the total number gas density, composed of purely hydrogen.  Since the wind is launched in the diffuse, ionized medium above the midplane, this is a good approximation. Of course, in the denser disk, near-complete ionization of gas not necessarily the case.  In these regions, the CR streaming speed boost determined with turbulent damping is overestimated as the CR streaming speed is proportional to the $n_i$, which will increase significantly. However, there are two reasons why our approximation is still justified.  First, the dense regions are not volume filling, so the impact of approximation is mitigated. Second, fast CR streaming likely occurs anyways in realistic scenarios, as other damping processes (i.e. ion-neutral damping) dominate in denser regions.


\section{Results}

\begin{figure*}
\centering
\includegraphics[width=16cm]{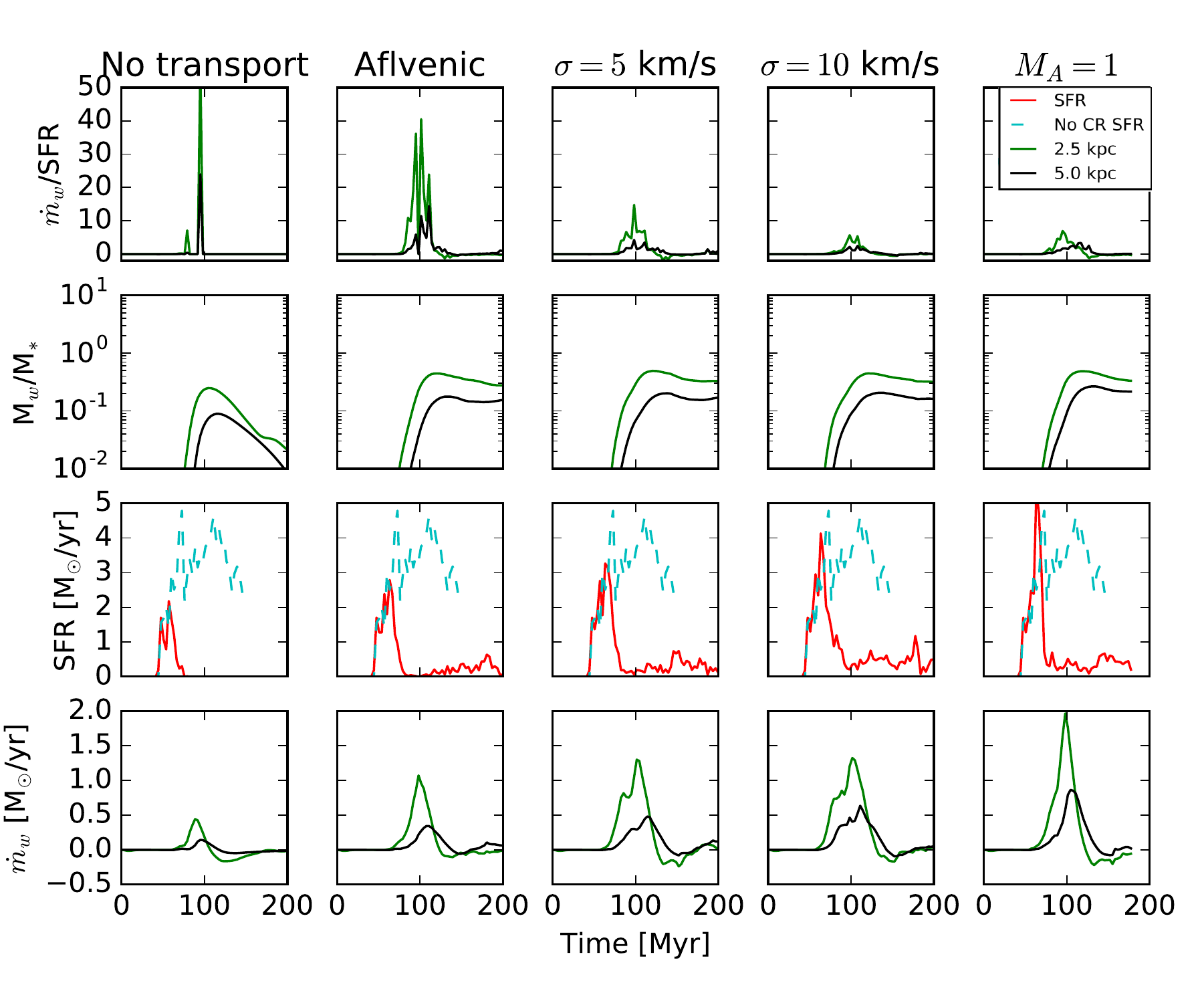}
\caption{Comparison of instantaneous mass loading (top row) through two surfaces above the mid-plane, integrated mass loading (second row), SFR (third row) and mass outflow (fourth row), between five different CR streaming implementations. The first column shows results from a simulation without CR transport, the second column shows results from simulations with Alfv\'enic streaming ($f = 1$), while the third and fourth show results from simulations with streaming including turbulent damping ($f = f(n_i,n_{\rm cr},B,\sigma = \textrm{ 5 km/s })$ and $f = f(n_i,n_{\rm cr},B,\sigma = \textrm{ 10 km/s})$ respectively), and fifth column shows results for constant $M_A = 1$, where the damping rates from \ref{regimetable} equal those from \citet{farmergoldreich2004}.}
\label{sfr}
\end{figure*}

We compare five simulations that differ in the value of the CR steaming speed $u_s = f u_A$ (see columns in Figure \ref{sfr}). The first case includes CRs without transport processes beyond advection with gas ($f = 0$). The second case is purely Alfv\'enic CR streaming with $f = 1$, where there is no dissipation of Alfv\'en waves.  The third and fourth cases correspond to locally-determined CR streaming following Eq. \ref{crss} using the appropriate turbulence regime for the environment.  The local streaming simulations assume a constant velocity dispersion, $\sigma =$ 5 km/s or 10 km/s, in the turbulent damping formulae from Table \ref{regimetable}. These values of velocity dispersion are on the order of the sound speed. The actual velocity dispersions reached at peak mass flow within the kpc wind launching region vary roughly between 10 to 25 km/s, which includes our assumed velocity dispersion of 10 km/s. Observationally, these values are plausible. For example, \citet{boettcher2016} found an upper limit of 25 km/s for turbulent velocity dispersion from optical emission-line spectroscopy in the edge-on galaxy NGC 891. The fifth case assumes constant  $M_A = 1$ (trans-Alfvenic turbulence) in the damping formulae, corresponding to the damping from \citet{farmergoldreich2004}.

%


\begin{figure}
\centering
\includegraphics[width=9cm]{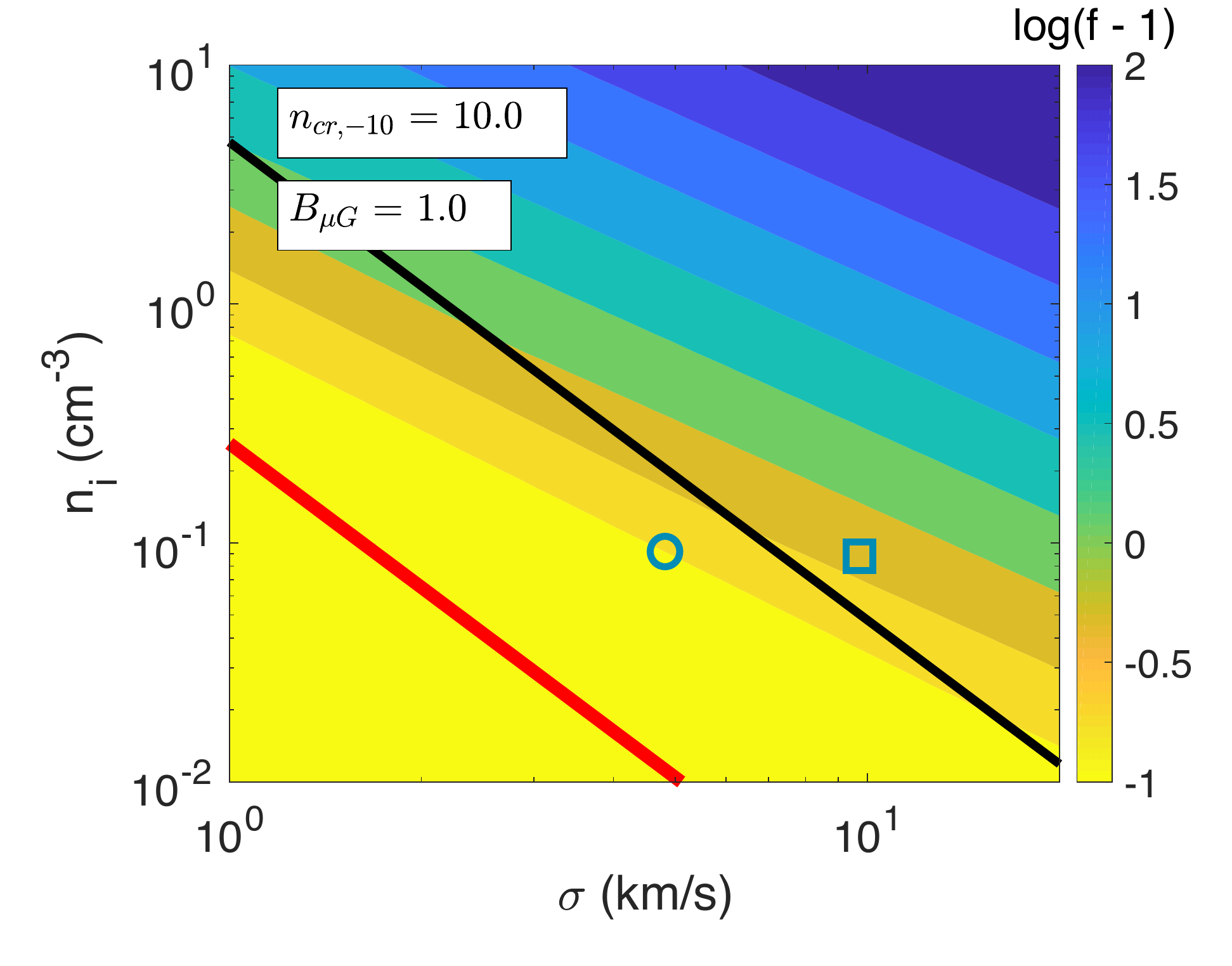}
\caption{The parameter space for the CR streaming speed boost $f-1$ factor for fixed magnetic field strength $B$ and CR number density $n_{\rm cr}$, and varying ion density $n_i$ and velocity dispersion $\sigma $. The magnetic field strength is shown in units of $\mu G$ and the CR number density is units of $10^{-10} cm^{-3}$. The damping formulae in Table \ref{regimetable} are used. Below a value of log($f-1$) = -1, shown in yellow, there is no significant super-Alv\'enic CR streaming and turbulent damping is ineffective. The black line indicates the boundary $M_{A} = 1.0$ and the red line indicates the transition from weak to strong, sub-Alfv\'enic turbulence. The parameters we use for the plot correspond to approximate values of quantities within the kpc size ionized ISM, above the thin galactic disk. The circle and box denote $n_i = 0.1 \rm{cm}^{-3}$ and velocity dispersions of 5 km/s and 10 km/s respectively. }
\label{parameter_space}
\end{figure}

\begin{figure*}
\centering
\includegraphics[width=17cm]{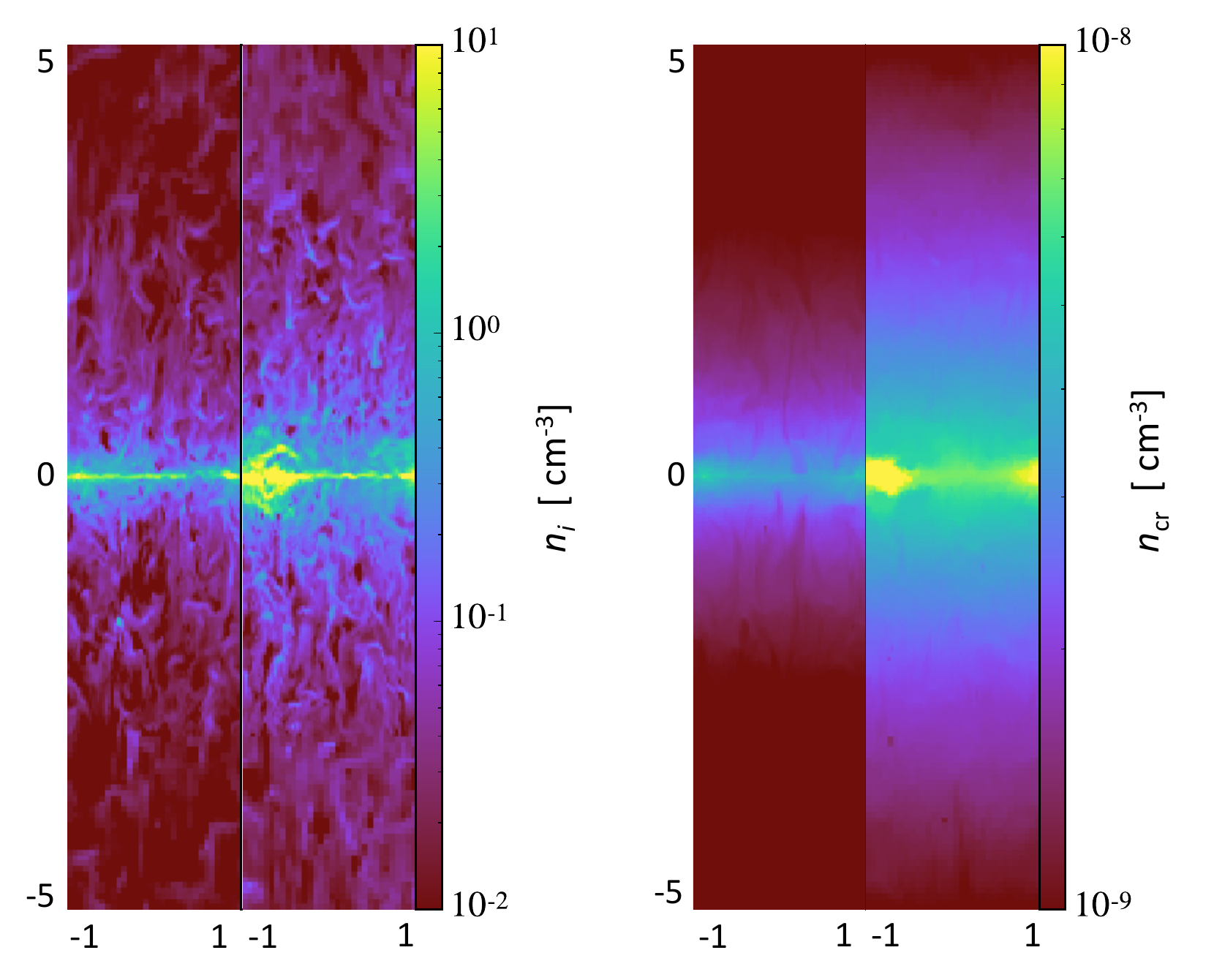}
\caption{Ion and CR number density slice of dimensions $\pm 5 $ kpc along  $z$ direction perpendicular to the midplane, for two simulations at 200 Myr: Alfv\'enic streaming and streaming including turbulent damping ($\sigma  = 10$ km/s). The Alfv\'enic simulation results are on the left hand side of each pair of plots, and the turbulent damping simulations are on the right side of each pair. The gas distribution (left pair) is slightly more extended in the turbulent damping simulation than for the Alfv\'enic streaming simulation. Similarly, the CR distribution (right pair) is significantly more extended in the turbulent damping simulation.}
\label{density}
\end{figure*}

Since the wind is launched \footnote{

We refer to the outflow as a `wind' even though the outflow velocity may not exceed the escape velocity of the halo. Cosmic rays continuously accelerate gas, so even if the velocity is not greater than the escape velocity near the disk, the gas can still reach the virial radius. Furthermore, in a global 3-dimensional gravitational potential, it is easier to launch a wind, compared to a one-dimensional potential, as the field strength decays faster \cite{martizzi2016}.}

from a roughly one kpc tall region just above the dense and thin galactic disk, we first examine where this region in our simulations falls in the parameter space of the CR streaming speed.  A discussion of the parameter space is important as we can only consider a limited number of simulations. Examining the parameter space allows us to develop intuition for the possible results of a simulation. Average values of the ionized ISM in our simulations are $n_{\rm cr} \sim 10^{-9} \textrm{cm}^{-3}, \ B \sim \mu G$, and $n_i \sim 10^{-1} \textrm{cm}^{-3}$. Figure \ref{parameter_space} shows the parameter space for these values, where CR streaming is expected to be nearly Alfv\'enic with log($f-1$) < 0.1, and also shows the transition to significantly super-Alfv\'enic streaming. 
At a velocity dispersion $\sigma = $ 10 km/s, the parameter space plot indicates that super-Alfv\'enic streaming is significant for the typical density of $n_i \sim 10^{-1} \textrm{cm}^{-3}$ in the launching region, as labeled by a blue box in the figure. Furthermore, since the typical ion density value we see is an average over all of the cells at a given height above the mid-plane, we can expect cells of even faster or slower CR streaming compared to $u_A$ from the slightly over or under-dense regions (see the slice plots in Figure \ref{density} for a qualitative look at the variation in density). Additionally, for the parameters shown, the wind launching regions is close to the black $M_A = 1$ line where turbulence transitions from $M_A < 1$ to $M_A > 1$. While we expect $M_A < 1$ farther out in the halo, closer to the disk the properties of turbulence will vary depending on the local conditions. In particular, an increase in the magnetic field strength will shift the $M_A = 1$ line to higher densities. This indicates that in the higher turbulence simulation, there will be regions where turbulence is super-Alfv\'enic because of lower magnetic field strength or higher ion density than average. At a velocity dispersion of $\sigma = $ 5 km/s, the typical gas density falls mostly in the region of Alfv\'enic streaming, as labeled by a blue circle in the figure.  However, the average density is large enough that slightly over-dense cells can still move into the super-Alfv\'enic regime, indicating that turbulent damping will still influence the global wind.

\begin{figure}
\centering
\includegraphics[width=8cm]{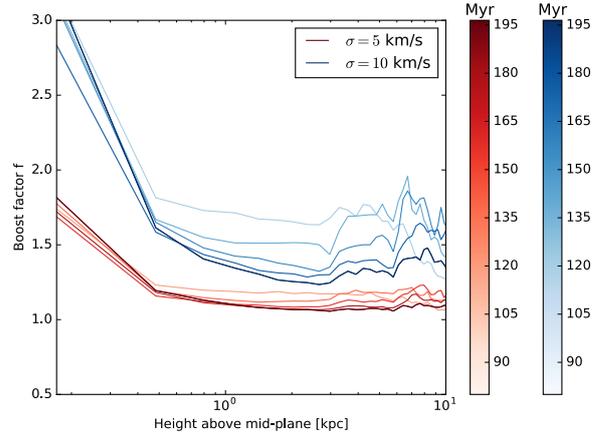}
\caption{Mass-weighted CR streaming speed parameter $f$ for local streaming with $\sigma = 5$ km/s (red) and $\sigma = 10$ km/s (blue).  The profile in the streaming speed boost factor $f$ is systematically higher for the stronger turbulence case, but both profiles approach near-Alfv\'enic streaming values away from the midplane. }
\label{fprof}
\end{figure}

Figure \ref{parameter_space} also illustrates the drastic effect that $M_A$ has on the CR streaming speed in the galaxy. The black line in the figure denotes the line at $M_A = 1.0$, where the turbulent damping formulae from \citet{lazarian2016} are equivalent to the \citet{farmergoldreich2004} damping formula for strong turbulence. The streaming speed at $M_A = 1.0$ given fixed $n_i,  n_{\textrm{cr}}$ and $B$ is always higher than the streaming speed for sub-Alfvenic turbulence as $\sigma$ decreases and other parameters are fixed. For example, on the $M_A = 1 $ line at $\sigma = 5$ km/s, the \citet{farmergoldreich2004} formula predicts a streaming speed boost log($f-1$) greater than 0.5, in the regime where turbulent damping is effective. If $M_A$ decreases, by lowering the velocity dispersion to $\sigma = 1$ km/s, at the same density, the streaming speed is essentially Alfv\'enic with log($f - 1)$ below a value of -1, where turbulent damping is not effective. All in all, not accounting for a decrease in $M_A$ results in an overestimation of the streaming speed. This effect is greater as $M_A$ decreases due to the steep dependence of the damping rates (and thus the streaming speed) on $M_A$, as shown in Table \ref{regimetable}. Such a decrease will change the average profile of streaming speed with height above the midplane in Figure \ref{fprof}. Furthermore, if the CR streaming speeds are different, the overall wind will be different, as will be shown in the subsequent part of the discussion regarding Figure \ref{sfr}.

%

%

\begin{figure*}
\centering
\includegraphics[width=16cm]{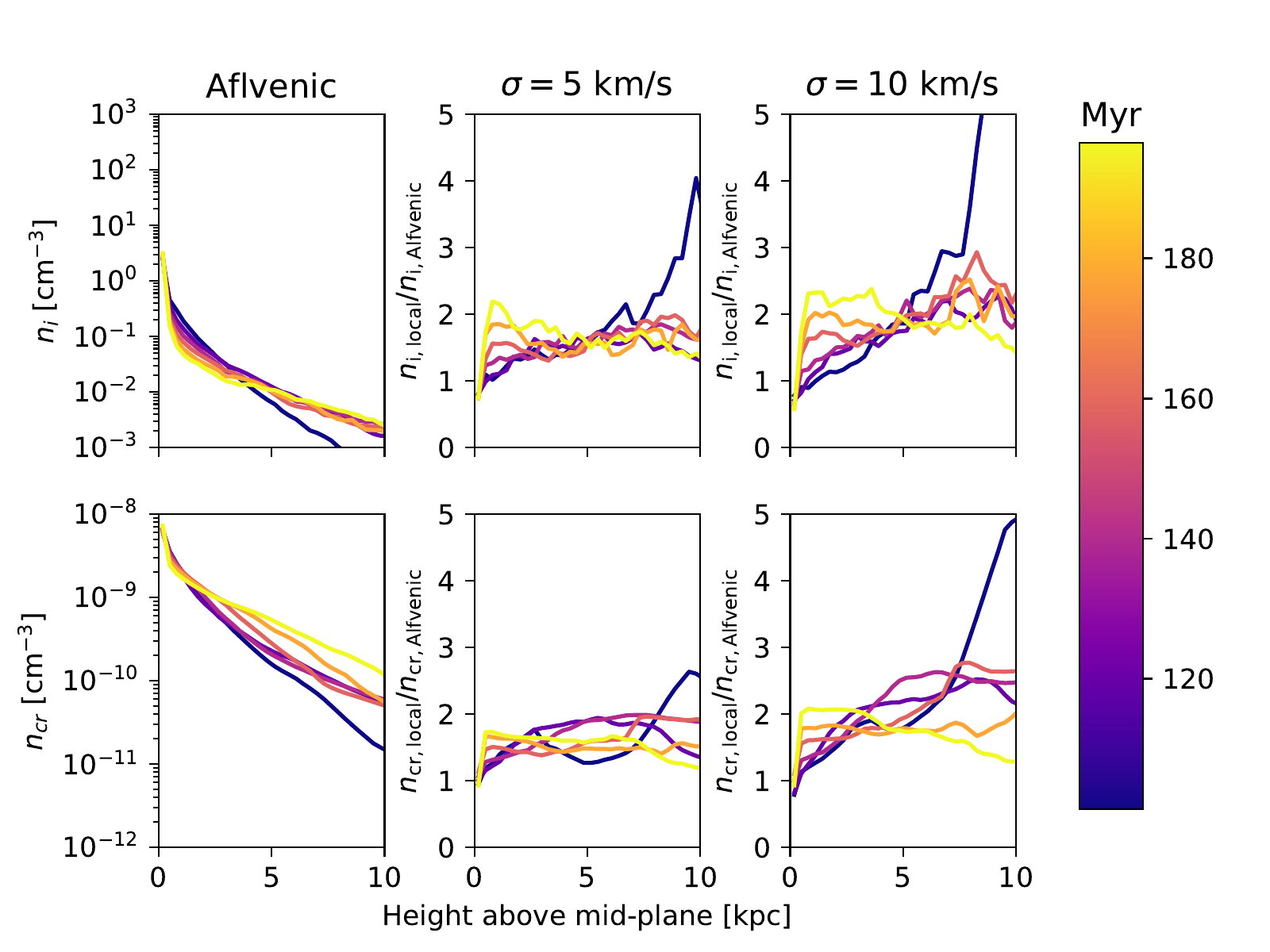}
\caption{Ion and CR number density volume-weighted profiles up to 200 Myr. Left column: Alfv\'enic streaming. Middle and Right columns: Profiles of the ratio of number densities found in different CR streaming simulations, turbulent damping for two strengths of turbulence, and the  Alfv\'enic streaming. The ratio is systematically greater than one for most of the evolution.}
\label{densprof}
\end{figure*}

Further comparisons we discuss here are between the Alfv\'enic and stronger turbulence ($\sigma = 10$ km/s) cases, as the weaker strength turbulence results follow similar trends to that of the stronger turbulence. Figure \ref{density} shows a qualitative comparison between the Alfv\'enic streaming and streaming with turbulent damping simulations in ion and CR density slices at a snapshot of 200 Myr, when the SFR and wind settles down. Figure \ref{densprof} shows the volume-weighted profiles of ion and CR density versus height above the mid-plane for the Alfv\'enic case, as well as the profiles for streaming with turbulent damping relative to the Alfv\'enic one. These profiles are volume-weighted to provide a fair comparison between the slice images and the averaged ion and CR densities. At 200 Myr there is a systematically more extended ion and CR distribution in the local turbulent damping case compared to the Alfv\'enic case. The slices show that the ion and CR densities is higher in the mid-plane for local streaming.  There is also a noticeable increase in the ion and CR densities farther away from the mid-plane at the top of the figure ($|z|$ = 5 kpc). The far right column of Figure \ref{densprof} quantifies these observations with relative profiles of ion and CR density. The general shape of the ion and CR density profiles with height is similar between simulations with Alfv\'enic streaming and those with damped streaming, as the relative density profile at later times in the simulation are roughly a constant factor different. The simulations with turbulent damping, however, have a systematically higher ion and CR density profile values at a given height compared with the Alfv\'enic streaming simulation. Within 5 kpc of the mid-plane, the average number density of both ions and CRs is about twice as large at a given height compared to the Afl\'enic streaming simulation. Additionally, at the mid-plane the CR number density profile approaches that of the Alfv\'enic streaming simulation. These results are due to the enhanced CR streaming near the mid-plane. With a boost in streaming, there is enhanced feedback--CRs can more effectively leave the dense mid-plane, allowing interaction with more tenuous gas and providing less pressure support against self-gravity of the midplane gas.  This leads to an increased SFR.

\begin{figure}
\centering
\includegraphics[width=8cm]{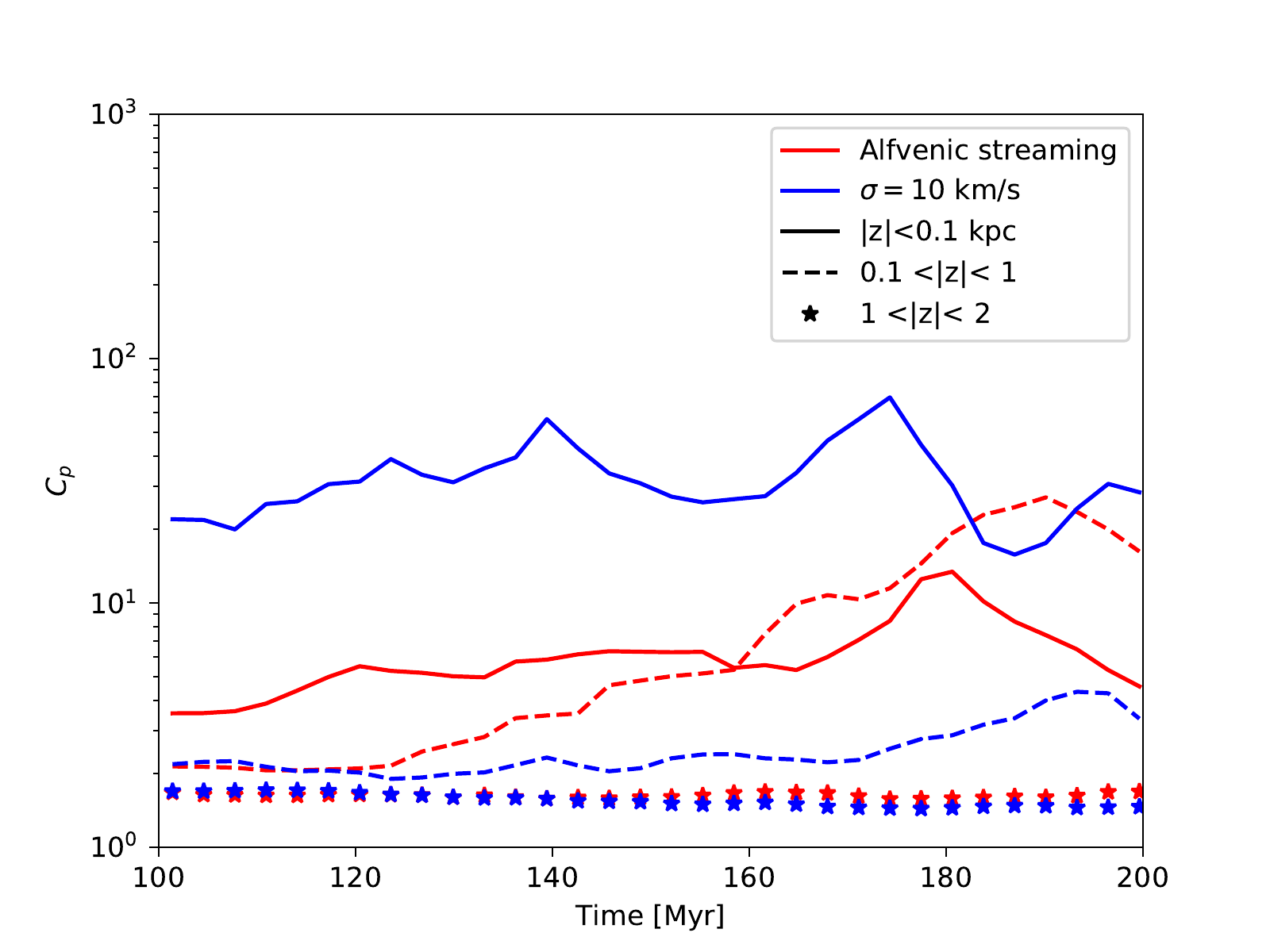}
\caption{Clumping factor of density $C_{p}$ for two simulations, Alfv\'enic (red) streaming and turbulent damping with $\sigma = 10$ km/s (blue), in three separate regions around the midplane. We focus on the second half of the simulation, from beginning of the wind at 100 Myr and the end of the simulation at 200 Myr. Close to the midplane ($|z| < 0.1$ kpc), clumping is relatively constant for both simulations, but there is stronger clumping for faster CR transport. In the intermediate region ($0.1$ kpc $< |z| <$ $1$ kpc), faster CR transport suppresses temporal changes in the clumping factor, leading to low values clumping. Farther out from the midplane ($1$ kpc $< |z| <$ $2$ kpc), clumping is low for both simulations. }
\label{clumpfig}
\end{figure}

In addition to denser gas on average, simulations with faster CR streaming also display differences in the relative clumping of the gas, which we quantify using the clumping factor
\begin{equation}
C_{p} = \frac{<\rho^2>}{<\rho>^2},
\label{clumpeq}
\end{equation}
where the averages are volume-weighted, as \citet{girichidis2018} calculates with simulations of a similar geometry as ours. Figure \ref{clumpfig} compares the average value of $C_{p}$ between simulations with Alfv\'enic CR streaming and the strongest turbulent damping case, within three regions: near the midplane ($|z| < 0.1$ kpc), intermediate height above the midplane ($0.1$ kpc $< |z| <$ $1$ kpc), and the lower halo ($1$ kpc $< |z| <$ $2$ kpc). Near the midplane, $C_p$ remains relatively constant throughout both of the simulations, with $C_{p} \approx 5-8$ in the Alfv\'enic streaming case and $C_{p} \approx 11-15$ for the turbulent damping case. The increased clumping in the turbulent damping case is consistent with the decreased stability in dense gas mentioned previously, as CRs more easily leave the midplane and the CR pressure support is reduced. In the intermediate height region, $C_{p}$ shows a strong temporal evolution in the Alv\'enic streaming simulation, starting relatively smooth with $C_{p} \approx 2$ and eventually reaching $C_{p} \approx 10$. The gas in this region is structured by supernovae \citep{girichidis2018}, consistent with the increased star formation rate (Figure \ref{sfr}) and clumping of gas with time. This increased clumping contrasts with the turbulent damping simulation, where $C_{p}$ begins at the same value as in the Alfv\'enic streaming simulation, and remains around $C_{p} \sim 2$ for the entire duration. The suppression of $C_{p}$ is due to the increased average density in this region relative to the Alfv\'enic streaming case (Figure \ref{densprof}), and that it is more difficult for supernovae to structure denser gas. In the lower halo up to 2 kpc in height, the gas remains consistently smooth in both simulations.

The boost in streaming speed is seen in the mass-weighted profile of CR streaming speed shown in Figure \ref{fprof}, where the stronger turbulence case is shown in blue. The profile is mass-weighted because we are interested in the influence that CRs have in accelerating gas. Near the mid-plane the boost factor $f$ is large and remains above $f=2$ within the thin galactic disk ($|z| < 500$ pc), and weakly super-Alfv\'enic at larger heights as turbulent damping becomes ineffective.

Since we do not track the CR spectrum or include energy dependent processes in our simulations, we cannot make predictions about observational signatures (e.g., synchrotron emission) produced by CRs.  Generally, a CR distribution more extended in height above the mid-plane does result in stronger radio emission in the halo around a galaxy \citep[][]{wiegert2015}. CR feedback could also influence the radio luminosity through its affect on the SFR \cite[e.g.][]{li2016}. Observational signatures derived from simulations with modeling of the CR spectrum \citep{yangruszkowski2017} will be investigated in future work.

The trends for the weaker turbulence strength follow those described for stronger turbulence, although, as expected, they are closer to the Alfv\'enic streaming results, due to the corresponding reduction in the strength of turbulence. Halving the velocity dispersion produces a significant reduction in streaming speed. For example, $f-1 \propto \sigma^{2}$ for strong, sub-Alfv\'enic turbulence, so halving $\sigma$ will reduce average streaming by a factor of four. Figure \ref{fprof} shows that both cases have high CR average streaming near the mid-plane and weakly super-Alfv\'enic streaming at larger heights, with the weaker turbulence run having a systematically smaller $f$ values.

The transition from no CR transport to progressively faster CR transport is seen in Figure \ref{sfr}. The figure shows a comparison for the SFR, mass outflow, and mass loading of five simulations: from left to right they are, `No transport' where $f = 0$, `Alfv\'enic' where $f = 1$, and finally three simulations with different turbulence strengths, two with constant velocity dispersion, and one with constant $M_A = 1$. The mass flux and integrated mass flux are calculated across two different surfaces (2.5 and 5.0 kpc) above the midplane.  The SFR plots include the result of a simulation without CRs. This case can also be thought of as the limiting case for maximal CR transport or equivalently no confinement of CRs.

For times up to 50 Myr, the simulations appear similar as there has not been significant and sustained CR production. After 70 Myr, there are enough CRs to influence feedback and the SFR profiles diverge. In the simulation without CRs, gas cools and collapses more easily to form stars due to the absence of CR pressure. For our specific implementation, this stellar feedback fails to launch a significant wind. The result is a comparatively greater peak SFR than in the other simulations with CRs. After the bulk of stellar particles form, the SFR in all of the simulations reaches an equilibrium around $0.5 \ M_{\odot}/$yr,  agreeing with the expected SFR of approximately $0.5 \ M_{\odot}/$yr for a gas surface density of $80 \ M_{\odot}/\rm{pc}^2$ \citep{kennicutt1998}.

In the case where there are CRs injected by feedback, but no streaming (i.e., $f = 0$ ), CRs are transported by advection with the gas. CRs cannot escape the midplane effectively, halting further cold gas gravitational collapse through additional pressure support that is not radiated away, unlike that of the gas. The first column of Figure \ref{sfr} shows the SFR and wind properties for a simulation with the CR transport being solely advection. There is a weak galactic wind, low SFR, and puffed-up disk morphology. These results agree with other works \citep[see][]{salembryan2014, girichidis2016, simpson2016, ruszkowski2017, farber2018}. 


When CR transport is added, CRs can escape the dense disk, influencing the evolution of the simulation. \citet{ruszkowski2017} included comparisons between no CR transport ($f = 0$) and Alfv\'enic CR transport ($f=1$), finding an increased SFR and stronger wind in the Alfv\'enic streaming case. They find the same trend of an increasing and more sustained SFR with progressively faster CR streaming, up to $f = 8$ or 8 times the Alfv\'en speed. Our simulations agree with this trend as the SFR increases for stronger turbulence, which leads to faster CR transport in our simulations. Our simulations extend their treatment of CR physics by allowing for spatial and temporal variations in the CR transport speed. The higher streaming speed in the ISM allows CRs to escape the dense mid-plane, allowing for further gas collapse.  Indeed, as the velocity dispersion increases, the SFR increases and approaches the peak SFR of a simulation without CRs. The differences in time evolution of the outward mass flux (bottom row in Figure \ref{sfr}) are weaker for increasing CR transport. The peak mass flux is similar for the simulations of Aflv\'enic streaming and both simulations with constant velocity dispersion; however, the mass flux does increase when turbulent damping is included in CR transport.  

Furthermore, the results in Figure \ref{sfr} confirm the idea, motivated by Figure \ref{parameter_space}, that accounting for the $M_A$ dependence of the damping formulae is essential. The fifth column of the figure shows the results for a simulation with constant $M_A = 1$ in the damping formulae. For sub-Alfv\'enic turbulence as is found in the wind launching region, at $M_A = 1$ the CR streaming speed is higher at a given $n_i$, $n_{\rm cr}$, and $B$ compared to the case where the velocity dispersion is constant.  At a given $n_i$ and $B$, $u_A$ is constant, so the velocity dispersion is a factor of $M_A$ different (smaller, for $M_A < 1$) than the equivalent value at $M_A = 1$. Correspondingly, the peak SFR is greater in the $M_A = 1$ case compared to the stronger velocity dispersion case of $\sigma = 10$ km/s, as CRs leave the midplane more quickly and allow more stars to form.  The peak mass flux is also much greater for the fifth simulation.



The increase in the SFR has a significant impact on instantaneous mass loading (ratio of the outward mass flux $\dot{m}_w$ to the SFR), seen in the top row of Figure \ref{sfr}. We find that the instantaneous mass loading is almost an order of magnitude smaller for the stronger turbulence simulation compared to Alfv\'enic streaming. This is because for stronger turbulence, CR transport is faster and the SFR is larger. This fact, combined with weaker sensitivity of the mass flux to CR transport speed, makes the instantaneous mass loading a decreasing function of the velocity dispersion. We plot the mass loading through two surfaces 2.5 kpc and 5 kpc above the mid-plane, finding similar trends. For the Alfv\'enic CR streaming simulation, the mass loading peaks at a value of about 40, while for the stronger turbulence simulation, it peaks at about 5. These values are consistent with observational constraints, which constrain instantaneous mass loading in range $10^{-2}$ and $10^1$ \citep{blandhawthorn2007}. For example, \cite{newman2012} use optical emission lines from a sample of galaxies and find mass loading factors of around 0.2-1 for the SFR surface density in our simulations.
Furthermore, in our simulations we only studied a single halo mass, a MW-like halo, rather than a variety of halo masses.
However, we expect that the feedback effects would be stronger for lower mass halos, and should follow trends similar to those we described above. \citet{jacob2018} studied a set of isolated galaxies (although CR transport was modeled with diffusion, rather than streaming as in this work) with halo masses ranging from $10^{10} M_{\odot}$ to $10^{13} M_{\odot}$ and found that mass loading scales like $M_{\rm{200}}^n$, with $-2 < n < -1$, so lower mass halos have larger mass loading. Mass loading values exceeding 10, as we find in the Alfv\'enic CR streaming case, can be suppressed by the presence of turbulence, landing within the observational constraints, as shown in the mass loading of the stronger turbulence simulation.

On the other hand, we find that the integrated mass loading (ratio of the total wind mass $M_w = \int_{0}^{200 \textrm{Myr}} \dot{m}_w$ d$t$ to the stellar mass $M_{\ast}$), is almost insensitive to the parameters we considered. The values we find for the integrated mass loading are about 0.2-0.3, roughly consistent with values from isolated slab simulations \citep[e.g.,][]{farber2018} and also global simulations \citep[e.g.,][]{ruszkowski2017}. While \citet{ruszkowski2017} found that the integrated mass loading increased with faster CR streaming, their simulations were global.

\subsection{Other relevant physical processes}
Our simulations do not account for various aspects of CR transport and CR interaction with the gas.

\begin{enumerate}

\item We focus on the streaming model of CR transport. CR transport is also modeled with diffusion \citep[e.g.][]{salembryan2014,girichidis2016, simpson2016}, with parallel and perpendicular diffusion coefficients with respect to the magnetic field, usually taken to be constant in space and time. Unlike in the streaming model, CRs do not heat the gas.  \citet{wiener2017} compared simulations with either the diffusion or streaming model of CR transport and
found that galactic winds are weaker with streaming compared to diffusion.  However, they did not include magnetic fields in their simulations, using the sound speed instead of the Alfv\'en speed for the streaming speed. The question of which model more accurately describes CR transport in the galaxy and halo remains without a definitive answer.

\item We do not include additional Alfv\'en wave damping processes such as ion-neutral damping and non-linear Landau damping. As ion-neutral damping is important when the medium that the CRs pass through is not completely ionized. The effects of ion-neutral damping on the galactic wind have been studied recently by \citet{farber2018}. Non-linear Landau damping occurs due to wave-particle interactions \citep{kulsrud2005}.  It is expected that this damping will not be dominant in astrophysical settings because it is self-regulating \citep[see][]{lazarian2016}.

\item Our implementation of turbulent damping of Alfv\'en waves from \citet{lazarian2016}, while still dependent on the local properties of the ISM and halo, is not completely self-consistent. First, we assumed a constant gas velocity dispersion in the turbulent damping rates, which strictly speaking is not true as turbulence will decay farther away from the star forming regions.  However, this assumption is sufficient in our case because the properties of the launching region (close to the mid-plane) should determine the properties of the overall wind.  Despite this assumption, the turbulent Alfv\'en Mach number (velocity dispersion divided by the Alfv\'en speed) nevertheless decays with height above the mid-plane as expected because the Alfv\'en speed increases with height.  The streaming speed profile in Figure \ref{fprof} also decays with height as expected, approaching Alfv\'enic streaming in the halo.  However, the assumption of constant velocity dispersion does not allow us to study in detail the different regimes of turbulence presented in \citet{lazarian2016}, as the halo turbulence strength (i.e. away from the wind launching zone) will be significantly overestimated. Second, we do not account for the level of ionization in the ISM and halo gas in determining the CR streaming speed. Only the ionized gas will participate in the growth and damping of Alfv\'en waves.  However, since the ISM and halo are significantly ionized, especially in the wind launching region just above the dense mid-plane, we did not specifically estimate the level of ionization in the gas and assume that the gas is fully ionized.

\item We also do not account for energy-dependent processes. The CR streaming speed boost $f-1$ increases with energy (or $\gamma_{\textrm{rel}}$), so higher energy CRs will escape faster. In a simple picture, we might expect that the CR spectrum will steepen from the initial spectrum. Furthermore, we do not account for loss of energy by CRs, due to hadronic losses as well as Coulomb losses, where CRs interact inelastically with atoms in the ISM.

\item There are also additional feedback mechanisms we did not implement. One example is the mechanism based on the radiation pressure from massive stars onto ISM dust in driving a galactic wind \citep{hopkins2012,zhangthompson2012}.  Including this mechanism involves calculating the radiation field using radiative transfer models, and including components of the ISM that interact with the field, such as dust, and is beyond the scope of this paper.

\item Finally, CR propagation depends on the details of the MHD turbulent cascade. In our model, we assume that turbulence only interacts with the CR-generated Alfv\'en waves and not the CRs themselves. In the extrinsic turbulence picture, turbulence scatters CRs as they propagate through the ISM with fast modes identified as the major agent of CR scattering \citep{yanlazarian2002}. A realistic magnetic topology is also important since CRs gyrate and follow the magnetic field and can also diffuse across field lines, potentially even super-diffusively \citep[i.e.,][]{lazarianyan2014}. We expect the galactic winds to be mostly sub-Alfv\'enic with the magnetic field not being strongly tangled, so the Alfv\'enic perturbations arising from galactic-scale turbulence will not scatter CRs efficiently \citep{yanlazarian2002,yanlazarian2004,yanlazarian2008}. Instead, the  Transient-Time damping (TTD) processes \cite[see][]{xulazarian2018} dominates the effect interactions of CRs with galactic-scale turbulence cascade. Overall, all of these processes interact in a non-linear fashion over a large range of scales, so a more complete description of CR transport in a galaxy remains to be fully understood.



\end{enumerate}

\section{Conclusions}

We perform three-dimensional magnetohydrodynamical simulations of a section of a galactic disk considering the impact of locally-determined cosmic ray (CR) transport on the properties of galactic winds. CR transport is treated within the self-confinement model, where the balance between wave growth and its decay by turbulent damping of self-excited Alfv\'en waves determines the bulk CR streaming speed relative to the gas. We employ the model of the streaming instability damping in \citet{lazarian2016} and find that the coupling of CRs experience significant spatial variations. Due to turbulent damping, the CRs are weakly coupled within the regions of the interstellar medium with higher level of sub-Alfvenic turbulence. We compared simulations with and without turbulent damping of the CR streaming instability. Our conclusions are as follows:

\begin{enumerate}

	\item We find that the star formation rate (SFR) increases when turbulent damping is included in the CR transport model and continues to increase with the strength of turbulence. Stronger turbulence damps confining Alfv\'en waves and leads to a corresponding boost in the average CR streaming speed.  As the CRs can leave the dense mid-plane more easily, the reduced pressure support from CRs allows gas to collapse and form stars more effectively.

	\item We show that the cumulative mass loading factor, the ratio of integrated wind mass to cumulative stellar mass, is insensitive to the impact of turbulent damping on the CR streaming speed.  For both strengths of turbulence tested, the cumulative mass loading factor asymptotes to the same value as the Alfv\'enic streaming run. 

	\item We show that the instantaneous mass loading is very sensitive to increased CR streaming speed due to turbulent damping.
       
	\item We demonstrate that the increased CR streaming speed due to turbulence results in more extended gas and CR density distributions. The larger SFR results in more stellar feedback, directly increasing the number of CRs produced in the mid-plane.  These CRs escape the dense mid-plane more quickly with an increased streaming speed, widening the CR distribution in height.  Escaping the central regions allows CRs to interact with lower density gas, which is easier to accelerate into the galactic wind, widening the gas distribution in height.
    
\end{enumerate}


\section*{Acknowledgements}
 We thank Ellen Zweibel for useful conversations. F.H. acknowledges support from the Rackham Merit Fellowship at the University of Michigan. M.R. acknowledges NASA ATP 12-ATP12-0017 grant and NSF grant AST 1715140. H.Y.K.Y.\ acknowledges support from NSF grant AST 1713722 and NASA ATP (grant number NNX17AK70G). AL acknowledges NSF AST graint 1816234 and NASA TCAN 144AAG1967. The software used in this work was developed in part by the DOE NNSA-ASC OASCR Flash Center at the University of Chicago. Computational resources for this work were provided by NASA High-End Computing Programing on the Pleiades machine at NASA Ames Research Center. Data analysis presented in this paper was performed with the publicly available \emph{yt} visualization software \citep{turk2011}. We are grateful to the \emph{yt} community and development team for their support.




\bibliographystyle{mnras}
\bibliography{cr_stream_turb_damp} 








\bsp	
\label{lastpage}
\end{document}